\documentclass[prd,twocolumn,aps,showpacs,nofootinbib,nobibnotes,superscriptaddress,preprintnumbers]{revtex4}
\usepackage{epsfig}
\usepackage{graphics}
\usepackage{bm}
\usepackage{color}
\usepackage{dcolumn}   
\usepackage{bm}     
\usepackage{bbm}       
\usepackage{amssymb}  
\usepackage{amsmath}
\usepackage{latexsym}
\usepackage{float}
\usepackage{ifthen}
\usepackage{caption,subfig}
\usepackage{enumerate}
\usepackage{url}
\usepackage{caption,subfig}
\usepackage{amsopn}

\usepackage{amsfonts}
\usepackage{multirow}
\usepackage{array}
\usepackage{booktabs}
\usepackage{rotating}

\usepackage{ulem}
\def\clap#1{\hbox to 0pt{\hss#1\hss}}

\def\({\left(}
\def\){\right)}
\def\[{\left[}
\def\]{\right]}
\def\bea{\begin{eqnarray}}
\def\eea{\end{eqnarray}}
\def\be{\begin{equation}}
\def\ee{\end{equation}}
\def\ba{\begin{eqnarray}}
\def\ea{\end{eqnarray}}
\def\beq{\begin{eqnarray}}
\def\eeq{\end{eqnarray}}
\def\mpl{M_{\rm P}}

\newcommand{\cs}{c_s}

\def\cs{c_{\rm s}}

\def\be{\begin{equation}}
\def\ee{\end{equation}}
\def\ba{\begin{eqnarray}}
\def\ea{\end{eqnarray}}
\def\beq{\begin{eqnarray}}
\def\eeq{\end{eqnarray}}

\def\mpl{M_{\rm P}}

\def\K{{\cal K}}

\def\L*{{\cal L}_*}
\def\L{\mathcal{L}}
\def\({\left(}
\def\){\right)}

\def\<{\langle}
\def\>{\rangle}

\def\cs2{c_{s}^{2}}

\def\be{\begin{equation}}
\def\ee{\end{equation}}
\def\ba{\begin{eqnarray}}
\def\ea{\end{eqnarray}}
\def\beq{\begin{eqnarray}}
\def\eeq{\end{eqnarray}}

\def\mpl{M_{\rm P}}

\def\K{{\cal K}}

\def\L*{{\cal L}_*}
\def\L{\mathcal{L}}
\def\({\left(}
\def\){\right)}

\def\<{\langle}
\def\>{\rangle}

\begin{document}

\title{Cosmology in doubly coupled massive gravity: constraints from SNIa, BAO and CMB}

\date{\today,~ $ $}

\author{Lavinia Heisenberg} \email{lavinia.heisenberg@eth-its.ethz.ch}
\affiliation{Institute for Theoretical Studies, ETH Zurich, 
\\ Clausiusstrasse 47, 8092 Zurich, Switzerland}

\author{Alexandre Refregier} \email{alexandre.refregier@phys.ethz.ch}
\affiliation{Institute for Astronomy, Department of Physics, ETH Zurich,\\
Wolfgang-Pauli-Strasse 27, 8093, Zurich, Switzerland}

\date{\today}

\begin{abstract}
Massive gravity in the presence of doubly coupled matter field via en effective composite metric yields an accelerated expansion of the universe.
It has been recently shown that the model admits stable de Sitter attractor solutions and could be used as a dark energy model. In this work, we
perform a first analysis of the constraints imposed by the SNIa, BAO and CMB data on the massive gravity model with the effective composite
metric and show that all the background observations are mutually compatible at the one sigma level with the model.
\end{abstract}


\maketitle

\section{Introduction}

Last year was the centenary year of Einstein`s General Theory of Relativity. 
This remarkable theory survived hundred years with great successes and
is still the fundamental theory that describes the underlying gravitational 
physics for a vast range of scales. Albeit a great deal of inquiry, it outlived
against most competitors of alternative theories. One of the first predictions
of the theory was the right amount of gravitational deflection of light, which
was confirmed by Arthur Eddington very soon after its inception. Nowadays 
the direct application of this in form of gravitational lensing is one of the
indispensable tools in astrophysics and cosmology. Another powerful 
prediction of General Relativity is the presence of gravitational waves.
These constitute the ripples of space-time itself, that travel outward from
a massive object in form of waves. Its discovery was a breathtaking
event \cite{Abbott:2016blz}.

Granting all this, there remains unsolved problems. Attempts to describe the
gravitational interactions by the principles of quantum mechanics failed tenaciously.
The absence of meaningful application of renormalizability techniques
diminishes the predictive power of the theory at large energy scales.
This problem has motivated to investigate ultraviolet modification of General
Relativity with the aim to successfully implement the quantum behaviour of
gravity. Furthermore, the presence of black hole and cosmological singularities 
are unwanted pathologies of the theory. It could be that the new physics of quantum
gravity automatically takes care of these singularities regularizing curvature
divergences. Alternatively, one could also consider classical modifications
in which curvature scalars are regular not due to quantum effects but rather
due to different behaviour of gravitational force at high energies implemented
in the modifications. This can also have important consequences for the
early universe allowing alternatives to the standard inflationary scenario \cite{Jimenez:2014fla,Jimenez:2015jqa}.

Another challenge was faced by the discovery of the accelerated expansion of the universe,
which is still one of the most intriguing problems in modern cosmology.
This detection has been now confirmed with high precision by many different
observations like Type Ia supernovae (SNIa), Baryon Acoustic Oscillations (BAO) 
and Cosmic Microwave Background (CMB) temperature power spectrum. 
Taking General Relativity as granted,
the expectation for the evolution of the universe would rather be a deceleration.
Therefore one is forced to inject some sort of unknown non-ordinary energy
into the theory. The inclusion of a cosmological constant indeed accounts for
most of the observations with very good precision requiring the value of the cosmological
constant to be of the order of $10^{-47}\text{GeV}^4$. The fact that it corresponds
to this very tiny value poses theoretical problems if one assumes that it corresponds
to the energy density of the vacuum of space, which is of the order of $\sim10^{120}$ larger.
This constitutes the worst problem of fine-tuning and is known as the cosmological constant
problem \cite{Weinberg:1988cp}. Even if one fine-tunes the value of the cosmological
constant to be very small at the classical level, this value is not radiatively stable. 
Considering quantum corrections in terms of matter loops will renormalize the value
of the cosmological constant proportional to the mass of the matter field and hence one
has to fine-tune the value at each loop order. This renders the theory unnatural. 
Similarly to the ultraviolet modifications to cure the pathologies at high energies, 
one can consider infrared modifications that could either tackle the old cosmological constant problem 
or provide a mechanism that accounts for the right dark energy phenomenology. 

In order to fit observations there is also the need for cold dark matter. Its origin
is still a mystery as well and has not been detected yet despite many efforts.
Together with the cosmological constant it builds the standard model of cosmology,
the $\Lambda$-CDM model. It prevails against all the alternative
models and explains for example perfectly well the observed fluctuations 
of the CMB and the structures on large scales. Despite the great agreement with 
the observations, some reported anomalies 
call for attention, even though they are statistically not very significant yet. Moreover, 
the model might have problems to account for the right observations of dark matter
at galactic scales and below, for example it fails to describe the tight correlations between 
dark and luminous matter in galaxy halos~\cite{SandMcG02,FamMcG12}.
This might be due to the lack of a complete understanding of the astrophysical phenomenology.
However, in this respect modifications in form of modified newtonian dynamics 
has been pursued ~\cite{Milg1} even though its successful extrapolation to
cosmological scales is problematic and calls for a better implementation of the
theory into a some sort of hybrid model \cite{Bekenstein:2004ne,Blanchet:2015sra,Blanchet:2015bia,Berezhiani:2015bqa}.

To address the above mentioned challenges one can consider modifications of
gravity in form of scalar-tensor~\cite{Horndeski:1974wa,Nicolis:2008in,Deffayet:2009wt,
Deffayet:2009mn,deRham:2011by,Burrage:2011cr,Heisenberg:2014kea}, 
vector-tensor~\cite{Horndeski:1976gi,EPU10,Jimenez:2009py,Jimenez:2013qsa,Heisenberg:2014rta,
Tasinato:2014eka,Allys:2015sht,Jimenez:2016isa} or tensor-tensor theories \cite{deRham:2010kj,Hassan12a}.
As a concrete infrared modification of General Relativity, the framework of
massive gravity has witnessed promising developments \cite{deRham:2010kj,Hassan:2011hr}, that could either be used
for dark energy \cite{deRham:2011by} or for the cosmological constant problem \cite{deRham:2010tw}. The theory requires the mass
of the graviton to be small but this small value is technically natural in the sense
that it is stable under quantum corrections \cite{deRham:2012ew,deRham:2013qqa,Heisenberg:2014rka}. Within possible scenarios the formulation
of the model in the presence of doubly coupled matter field via an effective composite
metric yields interesting cosmological solutions with stable perturbations \cite{deRham:2014naa,Gumrukcuoglu:2014xba}. 

In a previous 
work we have shown the presence of stable de Sitter attractor solutions \cite{Heisenberg:2016spl}. The presence of
de Sitter attractor does not guarantee a good fit to observations. We would like to
test the model using background observations like SNIa and distance priors. We first
introduce the framework that we are considering here in section \ref{sec:MG_eff} and
state the background equations of motion. For our analysis, the important modification
is encoded in the Hubble function which we compute in section \ref{sec:mod_Hub} by expressing
the equations of motion in terms of redshift before integrating them. After this preliminary 
analysis we first compare the modified Hubble function with the supernova data in section \ref{sec:SNI}
and put constraints on the model parameters. We further include the constraints coming from the
BAO data in section \ref{sec:BAO} and from the CMB data in section \ref{sec:CMB}. Finally, 
we compare the combined constraints of the model parameters in section \ref{sec:conclusion} and
show that all these background observations are mutually compatible at the one sigma level with our model.
\section{Massive gravity with effective composite metric}\label{sec:MG_eff}

The model that we would like to compare with background observations is the doubly coupled 
massive gravity model proposed in~\cite{deRham:2014naa}, where a matter field of the dark
sector is coupled to an effective composite metric built out of the dynamical and fiducial metric,
whereas the standard matter fields couple only to the dynamical metric. The cosmological
consequences of this model was already discussed in \cite{deRham:2014naa,Solomon:2014iwa,Gumrukcuoglu:2014xba}.
In a previous work we have further showed the presence of stable de Sitter attractor solutions,
making the model viable for dark energy studies. The action of the model reads
\begin{align}
\mathcal{S} = \int \mathrm{d}^4x \big[ \frac{\mpl^2}{2} \sqrt{-g}\left(R[g]-\frac{m^2}{2}\sum_{n=2}^4 \alpha_n{\cal U}[\cal K]  \right)  \nonumber\\
+\mathcal{L}^{\rm eff}_{\rm matter}(g_{\rm eff},\tilde{\rho},\tilde{P},\tilde{c}_s^2)+\mathcal{L}_{\rm matter}(g,\rho,P,c_s^2)\big]\,,
\label{lagrangian}
\end{align}
with $R[g]$ being the Ricci scalar of the dynamical metric and ${\cal U}[\cal K]$ denoting
the allowed ghost-free potential interactions between the two metrics $g$ and $f$ \cite{deRham:2010ik,deRham:2010kj}
\begin{align}
\mathcal{U}_2[\mathcal{K}] &=  2\left( [\K]^2-[\K^2]\right), \nonumber\\
\mathcal{U}_3[\mathcal{K}] &=[\K]^3-3[\K][\K^2]+2[\K^3],  \nonumber\\
\mathcal{U}_4[\mathcal{K}] &= [\K]^4-6[\K]^2[\K^2]+3[\K^2]^2+8[\K][\K^3]-6[\K^4]\,,
\end{align}
with $[..]$ denoting the trace and we already neglected the tadpole $\mathcal{U}_1$ and cosmological constant $\mathcal{U}_0$ contributions.
The effective composite metric is defined as~\cite{deRham:2014naa}
\begin{equation}
g^{\rm eff}_{\mu\nu} \equiv \alpha^2 g_{\mu\nu}+2\,\alpha\,\beta\, g_{\alpha\mu} \left(\sqrt{g^{-1}f}\right)^\alpha_{\nu} + \beta^2 f_{\mu\nu}\,,
\label{eq:geff}
\end{equation}
with the two arbitrary free parameters $\alpha$ and $\beta$. Actually, without loss of
generality one can fix $\alpha=1$ since the interesting dependence will be in the form of the ratio between the two parameters. This effective composite metric is special in the sense that its volume element corresponds
to the right potential interactions
\begin{equation}
\label{eq:detgeff2}
\sqrt{-g_{\rm eff}}= \sqrt{-g}\  \sum_{n=0}^4 \frac{(-\beta)^n}{n!}(\alpha+\beta)^{4-n} \mathcal{U}_n[K]\,.
\end{equation}
For the matter field of the dark sector that couples minimally to the effective composite metric we will assume a fluid with
energy density $\tilde{\rho}$, pressure $\tilde{P}$ and sound speed $\tilde{c}_s^2$ encoded in 
$\mathcal{L}^{\rm eff}_{\rm matter}(g_{\rm eff},\tilde{\rho},\tilde{P},\tilde{c}_s^2)$.
Note that its pressure can be very small but the important
requirement is that it is non-vanishing. And for the standard matter fields we will assume dust and radiation
type of matter fields that live on the dynamical metric represented by $\mathcal{L}_{\rm matter}(g,\rho,P,c_s^2)$.
We shall assume that the dynamical metric is of the form of the homogeneous and isotropic flat FLRW metric
$ds_g^2=-N^2 dt^2 +a^2 \delta_{ij} dx^idx^j$ and similarly the fiducial metric as 
$ds_f^2= f_{\mu\nu}dx^\mu dx^\nu = -\dot{f}^2dt^2 + a_0^2 \delta_{ij} dx^idx^j$.
Hence the effective metric is simply $ds^2_{\rm eff} = -N^2_{\rm eff} dt^2+a_{\rm eff}^2 \delta_{ij}dx^idx^j$,
with $N_{\rm eff} \equiv \alpha\,N+\beta\,\dot{f}$ and $a_{\rm eff} \equiv \alpha\,a+\beta\,a_0$ being the effective 
lapse and scale factor respectively. The modified Friedmann equation in this model corresponds to
\begin{equation}\label{eq:modFried}
3\,\frac{H^2}{N^2} =  m^2 \rho_A +\frac{\rho}{\mpl^2}+\frac{\alpha\,a_{\rm eff}^3}{\mpl^2\,a^3}\tilde{\rho}\,,
\end{equation}
with the energy density of the standard matter field $\rho$, the energy density of the matter field that lives
on the effective composite metric $\tilde \rho$ and the dimensionless effective energy density from the mass term being
$\rho_A\equiv U(A)-\frac{A}{4}\, \partial_AU$ where $U(A)  \equiv 6\,\sum_{n=2}^4\,\alpha_n (1-A)^n$ and
$A$ stands for the ratio of the scale factors $A\equiv a_0/a$. The acceleration equation of the system reads
\begin{eqnarray}\label{eq:accel}
\frac{2\,\dot{H}}{N^2}=\frac{2H\dot{N}}{N^3}+ m^2\,J\,A\,(r-1)-\frac{\rho+P}{\mpl^2}\nonumber \\
 - \frac{\alpha\,a_{\rm eff}^3}{\mpl^2a^3}\left[
\tilde{\rho} + \frac{N_{\rm eff}/a_{\rm eff}}{N/a} \tilde{P} \right]\,,
\end{eqnarray}
where $J=\frac13\partial_A\rho_m$ and $r\equiv \frac{\dot{f}/a_0}{N/a}$. The matter fields living on the dynamical and the effective composite metric have their corresponding conservation equations
\begin{eqnarray}
\frac{1}{N_{\rm eff}}\,\dot{\tilde{\rho}}+3\,\frac{H_{\rm eff}}{N_{\rm eff}}\,(\tilde{\rho}+\tilde{P})=0\,,  \nonumber\\
\frac{1}{N}\,\dot{\rho}+3\,\frac{H}{N}\,(\rho+P)=0\,.
\end{eqnarray}
Last but not least we have the Stueckelberg equation as constraint equation
 \begin{equation}
m^2\,\mpl^2J=\frac{\alpha\beta\,a_{\rm eff}^2}{a^2} \tilde{P}\,.
\label{eq:eqf}
\end{equation}
For the purpose of our present work, it will be convenient to express the background equations in terms of 
the redshift. In the next section we shall bring the relevant equations in
the form that will be most suitable for data comparison. Furthermore, we will assume $N=1$.
\section{Modified hubble function}\label{sec:mod_Hub}
We will first solve the constraint equation (\ref{eq:eqf}) for the pressure of the matter field in the dark sector
\begin{equation}
\tilde{P}= \frac{m^2 \mpl^2 a^2 J}{\alpha \beta (\beta+\alpha a)^2},
\end{equation}
and use the Friedmann equation (\ref{eq:modFried}) to solve it for its energy density 
\begin{equation}
\tilde{\rho}=-\frac{a^3(\rho_m+\rho_r+\mpl^2(-3H^2+m^2\rho_A))}{\alpha(\beta+\alpha a)^3} \,,
\end{equation}
where $\rho_m$ and $\rho_r$ are the energy density of matter and radiation respectively, living
on the dynamical metric $g$. After using these two equations, the explicit dependence on the matter field in the dark sector disappears. 
Next, we plug in the expressions for $\tilde{\rho}$ and $\tilde{P}$ into the acceleration equation (\ref{eq:accel}),
which simplifies to
\begin{equation}
\dot{H}=\frac12 \left( -3H^2-\frac{P_m+P_r}{\mpl^2}+m^2\left( \frac{(\beta + \alpha a)J}{\beta a}+\rho_A\right) \right)
\end{equation}
with $P_m$ and $P_r$ being the pressure of the matter fields that live on the dynamical metric.
We replace all the time dependent variables by their corresponding expressions in redshift
and their time differentiation by $\dot{\mathcal{V}}=-(1+z)H(z)\frac{d}{dz}\mathcal{V}(z)$,
where $\mathcal{V}$ stands for all the time dependent variables like $\rho_A(t)$, $H(t)$, $\rho_r(t)$...etc.
For the matter field we assume zero pressure $P_m$=0 and solving the continuity equation
gives for its energy density to be $\rho_m=\Omega_m (1+z)^3$, with $\Omega_m$ being the density parameter of matter.
 For radiation we assume $P_r=\frac{1}{3}\rho_r$ and solving
its continuity equation gives this time $\rho_r=\Omega_r (1+z)^4$ with the corresponding density
parameter for radiation $\Omega_r$. Hence the acceleration equation
in redshift space becomes
\begin{eqnarray}
&&12\mpl^2\beta(1+z)H\frac{dH}{dz}=3m^2\mpl^2(\beta(\kappa_1(2-4z) \nonumber\\
&&-z(2+z)\kappa_2+2(\kappa_2+\kappa_3))) +2\Omega_r(1+z)^4\beta\nonumber\\
&&+2\alpha\beta(\kappa_1+(1+z)(\kappa_2+\kappa_3+z\kappa_3)+\frac{9\mpl^2}{\alpha} H^2)
\end{eqnarray}
where we have introduced the combinations of parameters
$\kappa_1=3(\alpha_2+\alpha_3)+\alpha_4$, $\kappa_2=-2 (\alpha_2+2\,\alpha_3+\alpha_4)$
and $\kappa_3 =\alpha_3+\alpha_4$ for convenience. We can simply integrate the above
equation and obtain the evolution of the Hubble function, which results in
\begin{eqnarray}\label{HubbleFunction}
H^2&=&\frac{1}{6\mpl^2\beta}(2\Omega_r\beta(1+z)^4+\mpl^2(-m^2(-3z\beta(2\kappa_1 \nonumber\\
&+&(2+z)\kappa_2)+2\beta\kappa_3+\alpha(2\kappa_1+3(1+z)(\kappa_2 \nonumber\\
&+&2(1+z)\kappa_3)))+6\beta(1+z)^3c_1) \,,
\end{eqnarray}
with $c_1$ being an integration constant. Furthermore, it will be convenient to introduce the normalization $\Omega_mh^2$,  $\Omega_rh^2$...etc. with $H(z=0)=100 h \text{km} s^{-1} \text{Mpc}^{-1}$. Note also that the density parameter for radiation contains the contribution of
photons as well as the relativistic neutrinos
\begin{equation}
\Omega_rh^2=\Omega_\gamma h^2(1+0.2271N_{\rm eff}) \,,
\end{equation}
where $\Omega_\gamma h^2=2.469\times 10^{-5}$ at the CMB temperature $T_{\rm CMB}=2.725$ K
and $N_{\rm eff}=3.04$ stands for the effective number of relativistic neutrino species. In the following
sections we shall compare our model with the background observations like SNIa, BAO and CMB.
Note that we will assume $\Omega_k=0$ throughout the paper. Furthermore, without loss of generality 
we shall put $\alpha=1$ but keep $\beta$ arbitrary. Also since the integration constant $c_1$ in (\ref{HubbleFunction}) will be put in
relation to $\Omega_m$ through the Friedmann equation, we will be marginalizing over $c_1$.
\section{Constraints from SNIa}\label{sec:SNI}
\begin{figure}[t!]
	\includegraphics[width=0.9\columnwidth]{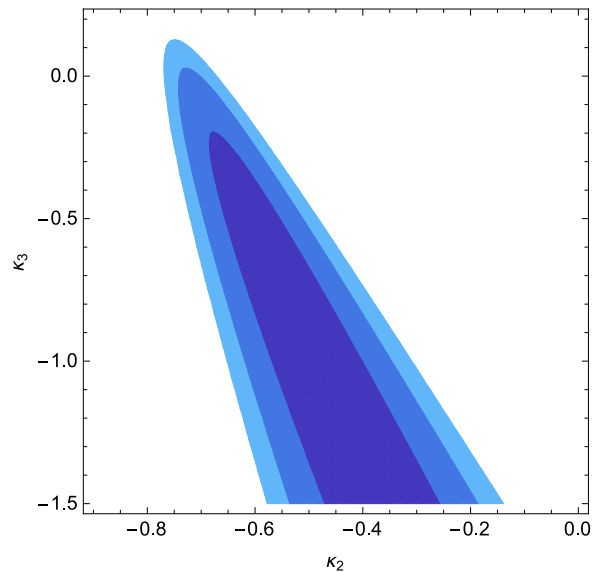}
\caption{We plot the marginalised $\chi^2$ estimator in the $\kappa_2$ and $\kappa_3$ parameter space. The $68\%$, $95\%$ and $99\%$ C.L. regions for the SNIa dataset union \cite{Kowalski:2008ez} are shown by the colour gradient. Recall that $\kappa_1=3(\alpha_2+\alpha_3)+\alpha_4$, $\kappa_2=-2 (\alpha_2+2\,\alpha_3+\alpha_4)$
and $\kappa_3 =\alpha_3+\alpha_4$, so this plot can be also seen as the constraints on the $\alpha_n$ parameters. We have chosen the $\kappa$ representation for convenience.}
\label{fig_SNIa}
\end{figure}

Supernova Type Ia are used as standard candles with known brightness to refer physical distances. The logarithm of the luminosity
of an astronomical object seen from a 10 parsecs distance gives its absolute magnitude, which on the other hand can be used
to give its brightness. We shall use the distance modulus $\mu$ to relate the expansion history of the universe to the apparent magnitude
of a supernova at a given redshift. It is defined as the difference between the apparent magnitude $m$ and
the absolute magnitude $M$ of the supernova and relates to the distance through
\begin{equation}
\mu=m-M=5\log D_L -5\log h +\mu_0 \,,
\end{equation}
with $\mu_0=42.38$ and dimensionless $D_L=H_0 d_L$. The luminosity distance on the other hand is given by 
$d_L=(1+z)r(z)$ with $r(z)$ standing for the comoving distance 
\begin{equation}
r(z)=\frac{1}{H_0}\int_0^z \frac{H_0}{H(\tilde{z})} d\tilde{z} \,.
\end{equation}
Once we have the distance modulus of our model, we can directly compare it with the supernova data
and compute the $\chi^2$ estimator
\begin{equation}
\chi^2_{\rm SN}=\sum_{i=1}^N\frac{(\mu(z_i;\beta,\kappa_1,\kappa_2,\kappa_3,h)-\mu_i)^2}{\sigma_i^2} \,.
\end{equation}
Since the supernova dataset is below redshift 2, we have neglected the contribution coming from
radiation, hence we have set $\Omega_r=0$ in this section. Since $h$ is degenerate with the absolute magnitude,
we marginalised over $h$. Furthermore, a careful analysis of the Likelihood reveals a degeneracy in
the $\beta$ parameter. A full detail scrutiny of the Likelihood would require a MCMC method applied to
the four dimensional parameter space, which is out of scope of the present analysis. We instead adapted to the grid-wise 
exploration of the Likelihood in the different directions and explored a local minimum. Due to the degeneracy in the
$\beta$ direction, without loss of generality we fixed it to a given number and the value of $\kappa_1$ and the integration constant $c_1$ to be the ones at the local  minimum for this given $\beta$ value, whereas marginalised over $h$. The local minimum of the Likelihood that we considered here is approximated by $\beta\sim 10$, $c_1\sim 0.27$ and $\kappa_1\sim -0.02$. Furthermore, $\Omega_m$ is related to the integration constant through the Friedmann equation. Thus out of the higher dimensional parameter space $\{ \Omega_m, \Omega_r, h, c_1, \kappa_1, \kappa_2, \kappa_3, \beta\}$, after marginalizing and/or fixing the minimum values for $\{ \Omega_m, \Omega_r, h, c_1, \kappa_1, \beta\}$ we were left with two parameters $\{ \kappa_2, \kappa_3\}$. Through the grid-wise exploration of the two dimensional Likelihood we constrained the remaining parameter space $\{\kappa_2,\kappa_3\}$ using the supernova data. We use the union data set \cite{Kowalski:2008ez}. The $68\%$, $95\%$ and $99\%$ C.L. regions for the supernova data is shown in Fig. \ref{fig_SNIa}. The local minimum of the $\chi^2$ estimator is around $(\kappa_2\sim-0.5, \kappa_3\sim-1.1)$. We could have chosen any other combination of pairs of the parameters in order to compare with the data, but since there is a degeneracy in the $\beta$ direction and the integration constant $c_1$ is related to the matter energy density, any combination of the $\kappa$ parameters could be equally good, but we could not explore further this possibility since this would require a full MCMC analysis of the model parameters. Our choice serves as a rule of principal how the supernova data can be nicely used to constrain the model parameters. 
\section{Constraints from BAO}\label{sec:BAO}
\begin{figure}[h!]
	\includegraphics[width=0.9\columnwidth]{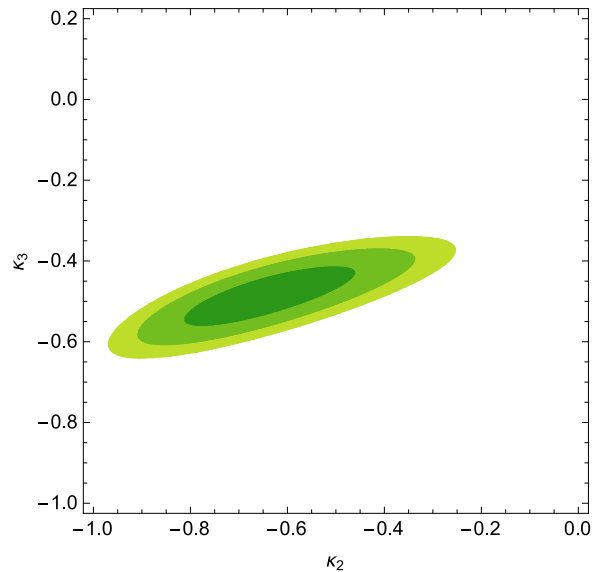}
\caption[]{This plot shows the $68\%$, $95\%$ and $99\%$ C.L. regions for the BAO. }
\label{fig_BAO}
\end{figure}

The density of baryonic matter has periodic fluctuations referred to as baryon acoustic oscillations, which
is the outcome of counteracting forces of pressure and gravity. The pressure released by the photons after
decoupling creates a shell of baryonic matter at the sound horizon. The measurement of these baryonic
oscillations yields the following distance-redshift relation at the redshifts $z=0.2$ and $z=0.35$ \cite{Percival:2007yw}
\begin{eqnarray}
\bold{V}_{\rm BAO} =
\begin{pmatrix}
\frac{r_s(z_d)}{D_V(0.2)} \\
\frac{r_s(z_d)}{D_V(0.35)}
 \end{pmatrix}= \begin{pmatrix}
0.1980\pm0.00588 \\
0.1094\pm0.0033
 \end{pmatrix} \,,
\end{eqnarray}
with the sound horizon expressed as
\begin{equation}
r_s(z)=\frac{1}{\sqrt{3}} \int_0^{\frac{1}{1+z}} \frac{da}{a^2 H(a)\sqrt{\left( 1+ \frac{3\Omega_b h^2}{4\Omega_\gamma h^2}a\right)}} \,,
\end{equation}
and the dilation scale as
\begin{equation}
D_V(z)=\left( r(z)^2\frac{z}{H}\right)^{1/3} \,.
\end{equation}
The redshift value $z_d$ represents the epoch at which the baryons were released from the photons and is
given by the fitting formula\cite{Eisenstein:1997ik}
\begin{equation}
z_d=\frac{1291(\Omega_mh^2)^{0.251}}{1+0.659(\Omega_mh^2)^{0.828}} \left( 1+b_1(\Omega_bh^2)^{b_2} \right) \,,
\end{equation}
with the parameters $b_1$ and $b_2$ standing for the short-cut notations
\begin{eqnarray}
b_1&=&0.313(\Omega_mh^2)^{-0.419}\left( 1+0.607(\Omega_mh^2)^{0.674} \right)  \\
b_2&=&0.238 (\Omega_mh^2)^{0.223}
\end{eqnarray}
The corresponding BAO data vector results in
\begin{eqnarray}
\bold{X}_{\rm BAO} =
\begin{pmatrix}
\frac{r_s(z_d)}{D_V(0.2)}-0.1980 \\
\frac{r_s(z_d)}{D_V(0.35)}-0.1094
 \end{pmatrix} \,,
\end{eqnarray}
and the $\chi^2$ estimator
\begin{equation}
\chi_{\rm BAO} ^2=\bold{X}_{\rm BAO}^T\bold{C}_{\rm BAO}^{-1}\bold{X}_{\rm BAO}
\end{equation}
with the inverse covariance matrix  \cite{Percival:2007yw}
\begin{eqnarray}
\bold{C}_{\rm BAO}^{-1} =
\begin{pmatrix}
35059 & -24031 \\
-24031& 108300
 \end{pmatrix} \,.
\end{eqnarray}
We are now ready to compare our model with the BAO data points. We can proceed in the same way as
for the SNIa data. However, note a crucial difference. The $\chi^2$ estimator for BAO depends directly on
the density parameter of the baryons $\Omega_b$, hence we need to marginalise over this parameter as well.
The parameters $\kappa_1$ and $\beta$ have been fixed to the value of the local minimum, whereas we
marginalised over $h$ again. The marginalised $\chi^2$ estimator over the parameters $(\kappa_2,\kappa_3)$ is given in Fig.\ref{fig_BAO}

\section{Constraints from CMB}\label{sec:CMB}
\begin{figure}[h!]
	\includegraphics[width=0.9\columnwidth]{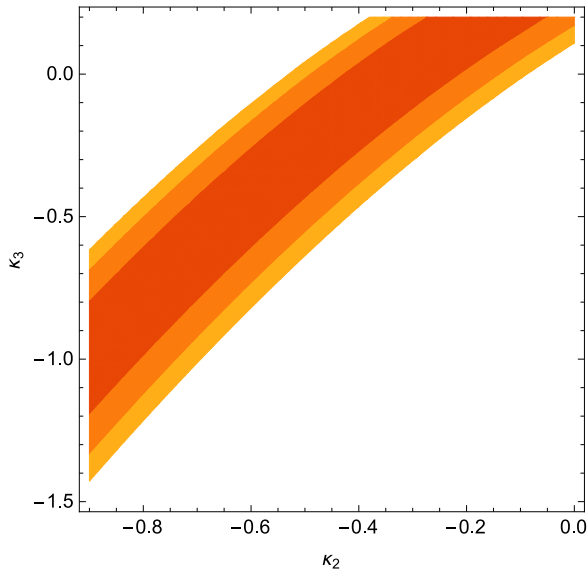}
\caption[]{This plot shows the $68\%$, $95\%$ and $99\%$ C.L. regions for the CMB. }
\label{fig_CMB}
\end{figure}

As next we would like to confront our model to the CMB data. For this purpose we tightly follow the distance priors
method of Komatsu et al \cite{Komatsu:2008hk}, which relies on the use of two distance ratios. The first distance ratio constitutes the ratio
between the angular diameter distance to the decoupling epoch and the comoving sound horizon size $r_s$ at the decoupling
epoch
\begin{equation}
l_A=\frac{\pi r(z_\star)}{r_s(z_\star)} \,,
\end{equation}
with the fitting function
\begin{equation}
z_\star=1048(1+0.00124(\Omega_bh^2)^{-0.738})(1+g_1(\Omega_mh^2)^{g_2})
\end{equation}
with the short-cut notations $g_1$ and $g_2$ standing for
\begin{eqnarray}
g_1&=&\frac{0.0783(\Omega_bh^2)^{-0.238}}{1+39.5(\Omega_bh^2)^{0.763}} \\
g_2&=&\frac{0.560}{1+21.1(\Omega_bh^2)^{1.81}}
\end{eqnarray}
The second distance ratio is the one between the angular diameter distance and the Hubble horizon size at the decoupling time
\begin{equation}
R=\sqrt{\Omega_mH_0^2}r(z_\star) \,.
\end{equation}
Following Kommatsu et al \cite{Komatsu:2008hk}, we take the following values for the distance priors
\begin{eqnarray}
\bold{V}_{\rm CMB} =
\begin{pmatrix}
l_A(z_\star) \\
R(z_\star) \\
z_\star
 \end{pmatrix}= \begin{pmatrix}
302.10\pm0.86 \\
1.710\pm0.019\\
1090.04\pm0.93
 \end{pmatrix} \,,
\end{eqnarray}
with the CMB data vector as
\begin{eqnarray}
\bold{X}_{\rm CMB} =
\begin{pmatrix}
l_A-302.10\\
R-1.710\\
z_\star-1090.04
 \end{pmatrix} \,,
\end{eqnarray}
and the inverse covariance matrix
\begin{eqnarray}
\bold{C}_{\rm CMB}^{-1} =
\begin{pmatrix}
1.800 & 27.968 & -1.103 \\
27.968& 5667.577 & -92.263 \\
-1.103 & -92.263 & 2.923
 \end{pmatrix} \,.
\end{eqnarray}
The corresponding $\chi^2$ estimator of the CMB is
\begin{equation}
\chi_{\rm CMB} ^2=\bold{X}_{\rm CMB}^T\bold{C}_{\rm CMB}^{-1}\bold{X}_{\rm CMB} \,.
\end{equation}
Note that these CMB distance priors are applicable only under the assumption that the
dark energy component is not relevant at the decoupling time. Since in our model of
massive gravity the modifications become appreciable only at small redshifts, the usage
of these priors is justified. 
In difference to the previous analysis, the CMB distance priors is very sensitive to the 
radiation component, hence we reinstal the explicit dependence of $\Omega_r$ in the
Hubble function. Furthermore, we have to marginalize over $\Omega_bh^2$ and $h$.
We show the $68\%$, $95\%$ and $99\%$ C.L. regions for the distance priors of
the CMB in Fig.\ref{fig_CMB}

\section{Conclusions} \label{sec:conclusion}
\begin{figure}[t!]
	\includegraphics[width=0.9\columnwidth]{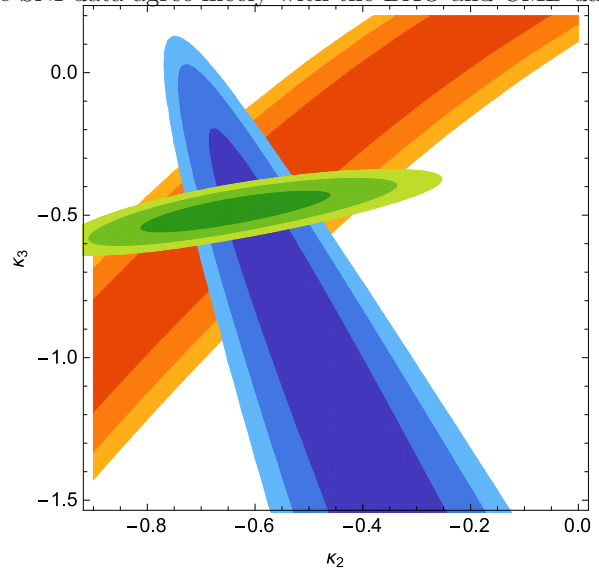}
\caption[]{This plot shows the $68\%$, $95\%$ and $99\%$ C.L. regions for the SNI, BAO and CMB data. We see nicely that the contours of the SNI data set is in nicely agreement with the BAO and CMB data set.}
\label{fig_all}
\end{figure}

This work was devoted to the detail study of the background evolution of massive gravity in the presence of
the composite effective metric to which the matter fields in the dark sector couple. Using the constraint and
Friedmann equation, we have seen that the direct dependence of the matter field in the dark sector disappears.
The resulting modified Hubble function only depends on the model parameters and the fluid dynamics
of the standard matter fields that live on the space-time metric. Clearly, in order to constrain the matter field
in the dark sector, one would need to go beyond background observations and consider the implications
coming in the perturbations. This shall be investigated in a future work. In a previous work, we had shown 
the existence of attractor de Sitter critical
points and studied the stability of perturbations. In this work, we have studied the constraints on the parameters
of the theory coming from SNI, BAO and CMB data. Since the model contains too many parameters,
we either fix or marginalised over some of the parameters leaving two parameters free. The two free parameters
appearing in the effective composite metric enter such that one of them can be fixed to unity. The remaining parameter
$\beta$ introduces degeneracy in the Likelihood. The model consists of six parameters $\{\Omega_m,h,\beta,\kappa_1,\kappa_2,\kappa_3\}$.
We first obtained the constraints coming from the SNI data. For this we explored grid-wise the Likelihood and
fixed the two parameters $\beta$ and $\kappa_1$ to the values of a local minimum whereas marginalised over $h$,
leaving the two parameters $\kappa_2$ and $\kappa_3$ free. In a similar way we compared our model to the
BAO and CMB data as well. As can be seen in
 Fig.\ref{fig_all}, the contours of the SNI data agree nicely with the BAO and CMB data even in the
 simple grid-wise exploration of the Likelihood of the parameters. The one sigma contours of the three observations
 overlap nicely and the preferred values for the two parameters are $\{\kappa_2\sim-0.6 ,\kappa_3\sim -0.5\}$, thus the 
 agreement is at the one sigma level. In a future work, 
 a detail analysis of the background observations will be performed using a MCMC method together with the 
 constraints analysis imposed by the observations coming from perturbations.

\acknowledgments We would like to thank J. Beltran Jimenez, R. Brandenberger, T. Kacprzak and S. Seehars for very useful and enlightening discussions. 
L.H. acknowledges financial support from Dr. Max R\"ossler, the Walter Haefner Foundation and the ETH Zurich Foundation.

\bibliography{cosmology_MG_geff_matter.bib}

\end{document}